\documentclass[nofootinbib,twocolumn]{revtex4}
\usepackage{graphicx}
\usepackage{amssymb}
\usepackage{amsmath}
\usepackage{bm}
\usepackage{hyperref}
\usepackage{soul}
\newcommand{\head}[1]{\textnormal{\textbf{#1}}}

\usepackage{dcolumn}   
\usepackage{bm,color}        

\newcommand{\vn}{\vec \nabla}

\newcommand{\be}{\begin{equation}}
\newcommand{\ee}{\end{equation}}

\newcommand{\R}{^{(3)}\!R}

\newcommand{\dd}{{\textrm d}}

\newcommand{\Lag}{{\mathcal{L}}}

\newcommand{\lsim}   {\mathrel{\mathop{\kern 0pt \rlap
  {\raise.2ex\hbox{$<$}}}
  \lower.9ex\hbox{\kern-.190em $\sim$}}}
\newcommand{\gsim}   {\mathrel{\mathop{\kern 0pt \rlap
  {\raise.2ex\hbox{$>$}}}
  \lower.9ex\hbox{\kern-.190em $\sim$}}}

\begin{document}



\title{Piercing the Vainshtein screen with anomalous gravitational wave
 speed: \\ Constraints on modified gravity from binary pulsars}

\author{Jose Beltr\'an Jim\'enez}
\affiliation{CPT, Aix Marseille Universit\'e, UMR 7332, 13288 Marseille,  France}

\author{Federico Piazza}
\affiliation{CPT, Aix Marseille Universit\'e, UMR 7332, 13288 Marseille,  France}

\author{Hermano Velten}
\affiliation{CPT, Aix Marseille Universit\'e, UMR 7332, 13288 Marseille,  France}



\begin{abstract} 
By using observations of the Hulse-Taylor pulsar we constrain the gravitational wave (GW) speed to the level of $10^{-2}$. We apply this result to scalar-tensor theories that generalize Galileon 4 and 5 models, which display anomalous propagation speed and coupling to matter for GWs. We argue that this effect survives conventional screening due to the persistence of a scalar field gradient inside virialized overdensities, which effectively ``pierces" the Vainshtein screening.  In specific branches of solutions, our result allows to directly constrain the cosmological couplings in the effective field theory of dark energy formalism.

\end{abstract}

\maketitle


{\bf Introduction - }Modifications of General Relativity (GR) that explain the  acceleration of the Universe can display a gravitational wave (GW) speed $c_T\neq 1$ (we use units $\hbar =c=1$). What are the observational constraints on this parameter? In some given model, $c_T$ can be expressed as a specific function of the (post-Newtonian) parameters of the theory, and thus constrained indirectly with solar system tests (see e.g.~\cite{Will:1993ns}). On the other hand, cosmological observations limit $c_T$ to the $10\%$ level (\emph{e.g.}~\cite{amendola}).
In Ref.~\cite{Moore:2001bv}, Moore and Nelson observe that subluminal GWs would be Cherenkov-radiated by particles traveling faster than $c_T$. By looking at high energy cosmic rays data, the authors manage to constrain this effect to the impressive level of $10^{-15}$. We notice, however, that the typical energy of the corresponding radiated gravitons, $\sim10^{10}$ GeV, is well above any reasonable cut-off of the modified gravity theories for cosmic acceleration. It is not difficult to envision \emph{e.g.} Goldstone modes in spontaneous Lorentz breaking situations that are subluminal at low frequencies and recover relativistic propagation above the symmetry breaking scale~\cite{NP}. With binary pulsars timing data, in this letter we obtain for $c_T$ looser limits ($\sim 10^{-2}$), which however  apply to frequencies that are relevant for an effective theory of dark energy.

One obvious objection is that scalar tensor theories generally come equipped with screening mechanisms, allowing to recover the stringent tests of gravity in the galaxy and in the solar system. Among these, the \emph{Vainshtein screening}~\cite{Vainshtein:1972sx} is particularly efficient, and relevant for those scalar tensor theories that display anomalous GWs speed.  
What screening guarantees, however, is  the suppression of the contribution of the scalar field $\phi$ to the total gravitational attraction between bodies in the Newtonian approximation. In a screened situation, the fluctuations of the metric field---gravitons---are left as the only mediators of long-range interactions. But not necessarily do they behave as in GR. The point is that the background value of the scalar $\phi_0$, although not directly participating in gravitational interactions, generally maintains a non vanishing gradient that spontaneously breaks Lorentz symmetry. In such a situation, 
 the effective gravitational Lagrangian need not be that of GR, even if it involves only massless gravitons. The Vainshtein screen is~\emph{pierced}.

The same mechanism  is responsible for other violations of the screening considered in the literature. In simple cases, deviations from GR boil down to a  spacetime variation of the Newton constant $G_N$. Refs.~\cite{Babichev:2011iz,Li:2013tda} use Lunar-laser-ranging to constrain this effect, obtaining limits on modified gravity models that are comparable in size to those obtained here. Preferred-frame effects~\cite{preferredframe} and  possibly anomalous values of the gravitational slip parameter $\gamma_{\rm PPN}$~\cite{Kimura:2011dc,Kobayashi:2014ida} (see also the following on this) have also been discussed in the literature. 

 {\bf Quadrupole formula, revisited - } For the sake of generality, we will consider a two-fold modification of GR encoded in the following  Lagrangian for the GWs sector:
\begin{equation} \label{theory}
{\cal L} \ =  \frac{1}{64 \pi G_{\rm gw}}\sum_{\alpha=+,\times}\left[\frac{1}{c_T^2}\dot \gamma_{\alpha}^2  -  \vert\vec{\nabla} \gamma_{\alpha}\vert^2 \right]\, ,
\end{equation}
where $+, \times$ represent the two polarizations of the GWs. 

First, we allow for a coupling of GWs to matter, $G_{\rm gw}$, possibly different than the Newton's constant $G_N$ inferred  in the Newtonian limit via the Poisson equation. Indeed, in addition to the radiating gravitons described by the above Lagrangian, we have the potential gravitons~\cite{walter},  responsible for the bound of the binary system. In modified gravity theories with an additional scalar degree of freedom, the scalar sector also becomes radiative. We will rely on Vainshtein screening while assuming that the contribution of the radiated scalar to the variation of the binary system period is negligible, as was shown to be the case in specific models~\cite{Galileonradiation}. 

The second modification that we consider is that GWs can propagate at a  speed $c_T$ different from the speed of light. We assume here that such a speed is constant,  direction- and polarization-independent. This statement is exact in the limit of a constant gradient for the background scalar field $\phi_0$, and in the reference frame where such a gradient is along the time direction. In the following we will quantify the corrections due to the presence of a spatial component of the gradient, and argue that in realistic situations such a component is negligible.

It is interesting to revisit, step by step, the standard derivation of the quadrupole formula (\emph{e.g.}~\cite{Maggiore:1900zz}) at the light of these modifications. First, we want to estimate the energy flux of a GW across a spherical surface at large distance $r$ from the source. The standard expression can be modified, essentially, 
by dimensional analysis (\emph{e.g.} by rescaling the time as  $\partial_{t}= c_T \partial_{t'}$ so that the GR formulae can be applied straightforwardly). We find
\begin{equation} \label{8}
\frac{\dd E}{\dd t}= \frac{r^2}{32 \pi c_TG_{\rm gw}}\int \dd\Omega \left\langle \partial_t \gamma_{ij} \partial_t \gamma_{ij}\right\rangle\, ,
\end{equation}
where $\langle \dots \rangle$ means average over a region of spacetime much larger than the GW wavelength. On the other hand, at the lowest (quadrupole) order in the velocity expansion, the radiated amplitude of GWs from a given source is obtained with the usual formulae, barring the replacement $G_N\rightarrow G_{\rm gw}$ and the different retarded time at which the source is evaluated, 
\begin{equation}
[\gamma_{ij}]_{quad} \ = \ \frac{2 G_{\rm gw}}{r}  \ddot{Q}_{ij}^{TT} \!\left(t - \frac{r}{c_T}\right),
\label{solgamma}
\end{equation}
where $Q_{ij}^{TT} $ is the transverse-traceless projection of the quadrupole moment $Q_{\ij}$ of the source. Note that $\ddot{Q}_{ij}^{TT}$ appears in (\ref{solgamma}) after using the energy-momentum conservation of matter. Since the matter sector has the usual Lorentz symmetry, the time derivatives acting on $Q_{ij}^{TT}$ do not introduce any additional factors of $c_T$. As in the standard calculation, the way such a projection is made depends on the direction of the GW and this should be taken into account when calculating the surface integral~\eqref{8}. This results in the following total power emitted
\begin{equation}
P_{quad}=\frac{ G_{\rm gw}}{5 c_T} \left\langle  \dddot{Q}_{ij} \dddot{Q}_{ij}\right\rangle.
\label{Pquad}
\end{equation}
This expression coincides with the formula obtained in \cite{Blas:2011zd} for Horava gravity.

{\bf Binary pulsar constraints - } By the above modified quadrupole formula, binary pulsars observations will allow us to constrain the combination $c_T G_{\rm gw}$, modulus some assumptions on the expressions of the Keplerian parameters of the bound system that we detail in the following. 
The emission of GWs results in a decrease of the orbital period $P_b$~\cite{Maggiore:1900zz}. {\it Mutatis mutandis}, we get
\begin{align} \label{Pdot}
\dot{P}_b =&-\left(\frac{G_{\rm gw}}{G_N}\frac{c}{c_T}\right) \frac{192 \pi G^{5/3}_N}{5 c^5}  \left(\frac{P_b}{2 \pi}\right)^{-\frac53}(1-e^2)^{-\frac72}\\ \nonumber
&\times\left(1+\frac{73 e^2}{24}+\frac{37 e^4}{96}\right) m_p m_c (m_p + m_c)^{-1/3},
\end{align}
where $e$ is the eccentricity of the Keplerian orbit and $m_p$ and $m_c$ are the masses of the pulsar and its companion, and we have temporarily reintroduced (just here and in~\eqref{new6}) the dimensional  
speed of light $c$. As explained below Eq. \eqref{theory}, we assume potential gravitons and radiative gravitons to couple to matter with different strengths. Note the different roles in the derivation played by $G_N$, coming from the formula of the orbits, and $G_{\rm gw}$, coming from the actual emission of gravitational waves.

We use the most accurate available data on $\dot P_b$, those of the Hulse-Taylor pulsar (PSR B1913+16)~\cite{Hulse1975}, with the orbital parameters shown in Table \ref{Table1}\footnote{Eq. (\ref{Pdot}) is calculated in the orbiting system reference frame which is accelerated with respect to the solar system barycenter frame \cite{Damour1991}. This effect,  known as Shklovskii effect, gives an extra $\Delta \dot{P}_{b,gal}=-0.027 \pm 0.005 \times 10^{-12}$ which should be subtracted.}. Before using this information, we need the standard expressions for the advance of the periastron $\dot{\omega}$ and the amplitude of the Einstein delay
$\gamma$ \cite{Will:2014xja}, which also depend on the Keplerian parameters $e$ and $P_b$, and on the masses $m_p$ and $m_c$. We can thus use the binary pulsar data to constrain the combination $c_T G_{\rm gw}$, in addition to the two masses. 

\begin{table}
\centering
\begin{tabular}{ccc}
  \hline
  \head{Parameter} &\head{Description} &\head{Value} \\
  \hline
  $e$ & eccentricity & 0.6171334(5) \\
  $P_b({\rm days})$ & period & 0.322997448911(4) \\
  $\dot{w} ({\rm deg/ yr})$  & periastron advance& $4.226598(5)$ \\
  $\gamma ({\rm ms})$ & Einstein delay & $4.2992(8)$ \\
	$\dot{P}_b$ & period decay& $-2.423(1) \times 10^{-12}$ \\
	\hline
\end{tabular}
\caption{Orbital parameters for PSR B1913+16 from \cite{Weisberg2010}.
} 
\label{Table1}
\end{table}

\begin{figure}[t]
\includegraphics[width=0.44\textwidth]{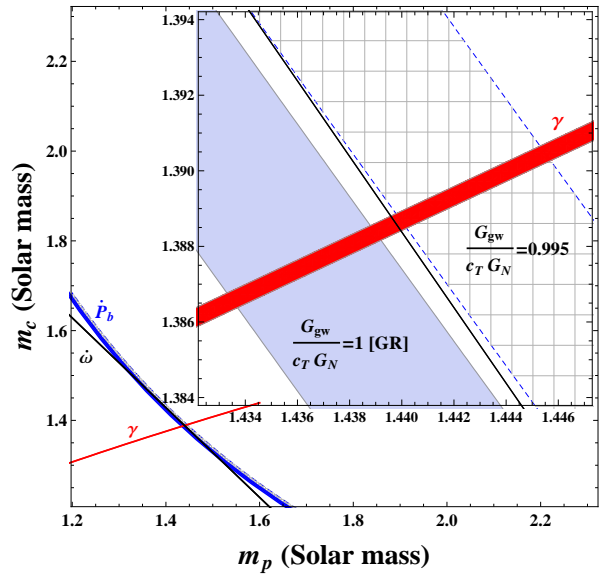}\\ 
\caption{
Mass-mass diagram for PSR B1913+16 (the Hulse-Taylor pulsar) based on the post-Keplerian parameters $\dot{w}$ (black), $\gamma$ (red) and $\dot{P}_b$ (blue). Varying the combination $c_T G_{{\rm gw}}/G_N$ amounts to shifting the 1-$\sigma$ stripe of $\dot{P}_b$. }
\label{fig1}
\end{figure}

While the expression of $\gamma$ is derived, essentially, in the Newtonian approximation, a comment regarding the parameter $\dot{\omega}$ is in order here. In a modified gravity set-up, such a quantity depends on both post-Newtonian parameters $\gamma_{\rm PPN}$ and  $\beta_{\rm PPN}$~\cite{Will:2014xja,Wex:2014nva}. Since we are aiming (see below) to a precision of $10^{-2}$, we  rely on solar system tests, which constrain $\gamma_{\rm PPN}$ and  $\beta_{\rm PPN}$ at the levels of $10^{-5}$ and $10^{-3}$ respectively, and $\dot{\omega}$ directly and independently, for Mercury, at the level of $10^{-3}$~\cite{cassini,Will:2014xja,Damour:1996xx}.

We can now proceed to construct the mass-mass diagram and the corresponding constraints on $c_T G_{\rm gw}$ as shown in Fig. \ref{fig1}. We  see  that the binary pulsar data meet in a small region of the $(m_p, m_c)$ plane. GR predictions fall in the intersection of $\dot \omega$ and $\gamma$ within about 1$\sigma$ confidence. By imposing compatibility of the three constraints at the 1$\sigma$ level we obtain the following bound:
\begin{equation} \label{new6}
0.995 \  \lesssim \  \frac{G_{\rm gw}}{G_N} \frac{c}{c_T} \ \lesssim \ 1.00\, .
\end{equation}

{\bf Symmetries and scalar field gradients -}  We are now to discuss the implications of the above bound on concrete scalar tensor models for dark energy. First, it is helpful to consider the basic structure of the simplest scenario that displays an anomalous GW speed, the \emph{quartic} galileon model, with Lagrangian 
\be \label{quartic}
\Lag_4^{\rm gal}=-\frac{X}{\Lambda^6}\Big[(\Box\phi)^2-(\nabla_\mu\nabla_\nu\phi)^2 \Big] \, .
\ee
In the above, $X\equiv\partial_\mu\phi\partial^\mu\phi$ and $\Lambda$ is some energy scale of the order of $\Lambda\simeq(M_p H_0^2)^{1/3}$ with $H_0$ the Hubble parameter today. By inspection of the second term inside the square brackets, we see that the covariant derivatives generate a term quadratic in the Christoffel symbols. In the presence of a background field $\phi_0$ with non-vanishing timelike gradient, such term contributes to the quadratic Lagrangian for the gravitons $h_{ij}$ as  $\sim \dot h_{ij}^2$, thus modifying the propagation speed of GW. It is immediate to see that $c_T$ is dependent only on the gradient of $\phi_0$ in this case. Only when $\nabla_\mu \phi_0$ vanishes does $c_T$  go to one.  

But for theories enjoying  \emph{shift symmetry} $\phi \rightarrow \phi + const.$, the actual value of the scalar is irrelevant, and there is no evident mechanism for it to detach from cosmic evolution and become constant inside a virialized object. This means that we do expect, in general, a non-vanishing scalar gradient inside screened environments---a ``local remnant" of the expansion of the Universe. 

Indeed, the profile of a cosmologically evolving scalar field in the presence of a matter source is easily estimated for those theories that enjoy a further, galileon, symmetry~\cite{NRT}, which makes a constant gradient of $\phi$, and not only its actual value, irrelevant. In Minkowski space, this is defined as the invariance under $\nabla_\mu \phi \rightarrow \nabla_\mu \phi + b_\mu$ with $b_\mu$ a constant vector. 
Let $\phi_0^{\rm cosm}(t)$ be the cosmological solution obtained under the assumption of homogeneity and isotropy. Well inside the Hubble radius, where the metric is similar to Minkowski, this is effectively a field configuration of constant gradient. Once we find a suitable radial solution $\phi_0^{\rm astro}(r)$ vanishing at infinity around some localized matter source, in virtue of galileon symmetry, we can simply add the two solutions,
\begin{equation} \label{split}
\phi_0(r,t) \ \simeq \ \phi_0^{\rm cosm}(t) + \phi_0^{\rm astro}(r)\, .
\end{equation}
Galileon theories are a combination of 5 Lagrangian terms with an increasing number of fields $\phi$.  

As a case study, let us consider the \emph{quartic} galileon~\eqref{quartic}. 
The cosmological gradient for this theory is given by $\dot{\phi}^{\rm cosm}_0\sim H_0M_p\sim \Lambda^3 H_0^{-1}$. On the other hand, the analysis of the Vainshtein mechanism near a spherically symmetric object shows that the radial gradient of the scalar inside the screened region for quartic Galileon is constant, $(\phi_0^{\rm astro})'\sim (M/M_p)^{1/3}\Lambda^2\sim r_V\Lambda^3$, where we have introduced the Vainshtein radius $r_V\simeq(M/M_p)^{1/3}\Lambda^{-1}$ and $M$ is the mass of the matter source. In summary, 
\be \label{ratio}
\frac{\phi'_0}{\dot{\phi}_0}\sim\frac{r_V}{H_0^{-1}}\, ,
\ee
which shows that a localised source contributes a very mild radial component to the total gradient of the field. For example, the Sun has $r_V \sim 1$ kpc so this ratio is of order $\sim 10^{-6}$.  In comparison, our peculiar velocity with respect to the CMB gives a much larger  (effect$\sim 10^{-3}$). Our estimates are in agreement with the explicit numerical calculations of~\cite{ferreira}.

Gravity inevitably breaks the symmetry $\nabla_\mu \phi \rightarrow \nabla_\mu \phi + b_\mu$, if anything, because there is no such thing as a constant vector $b_\mu$ in a general  spacetime. However, we can apply the above estimates to all scalar tensor theories that reduce to galileon in the \emph{decoupling limit}, formally defined as $M_P\rightarrow \infty$ while keeping $\Lambda$ constant.  Among these, theories with \emph{weakly broken galileon symmetry}~\cite{TV} have their Lagrangians protected against quantum corrections.

{\bf Cosmological EFT operators -} We have just shown that $\phi_0'\ll\dot\phi_0$ (even) inside the Vainshtein radius, where the non linearities in the scalar can become important but the metric is very close to Minkowski. The most general quadratic Lagrangian for the metric fluctuations in the presence of a background scalar field of constant timelike gradient
is conveniently studied within the effective field theory (EFT) formalism for cosmological perturbations~\cite{GPV,GLPV,PV}. 
By choosing the time coordinate to be proportional to the scalar field (unitary gauge), all degrees of freedom are transferred to the metric, chosen to be the one minimally coupled to matter (\emph{Jordan frame}). A limited number of operators capture the linear dynamics of the most general scalar-tensor theory with an equation of motion of at most second order for the propagating scalar fluctuation~\cite{GLPV}. Among such operators, only three affect the pure graviton sector,  
\begin{equation} 
{\cal L}  \supset  \frac{M^2}{2} \left[R  + \epsilon_4 \left(\delta K^{i j} \delta K_{ij} -  \delta K^2  \right)   - \tilde \epsilon_4 \,  \R \delta N  \right]\, ,
  \label{example}
\end{equation}
where $R$ is the Ricci scalar, $\delta K^{ij}$ is the perturbation of the extrinsic curvature $K^{ij}$ of the $t = const.$ hypersurfaces, $\R$ their Ricci scalar and $\delta N$ the perturbation of the lapse function. $M$, $\epsilon_4$ and $\tilde \epsilon_4$ are time dependent coefficients. In GR, $M=const.$, $\epsilon_4 = \tilde \epsilon_4=0$. 
The above operators arise \emph{e.g.} in the class of models introduced  in~\cite{G3} as a generalization of \emph{Horndeski theory}, which is the most general scalar-tensor theory with equations of motion of at most second order~\cite{horndeski,Deffayet:2009mn}. We refer the reader to~\cite{G32} for the expressions of $\epsilon_4$ and $\tilde \epsilon_4$ as functions of the full \emph{Beyond Horndeski} Lagrangians.

To study the effects of the terms~\eqref{example} it is convenient to switch to Newtonian gauge on a Minkowski background. By forcing a time-diffeomorfism $t\rightarrow t+ \pi$, the fluctuations of the scalar field $\pi$ reappear in the action, after which we can fix the metric to have the form  
\begin{equation} \label{newtonian}
\dd s^2 = - (1+2\Phi) \dd t^2 +\left[(1-2 \Psi)\delta_{ij} + \gamma_{ij}\right]\dd x^i \dd x^j\, ,
\end{equation}
where $\Phi$ and $\Psi$ are the two Newtonian potentials and $\gamma_{ij}$ represents the transverse traceless graviton.
At highest order in derivatives the quadratic Lagrangian reads~\cite{G32}
\begin{align}
{\cal L}\, =& \, \frac12  g^{\mu \nu} T_{\mu \nu} + M^2 \left[\frac{1}{4c_T^2}\left(\dot \gamma_{ij}^2 - c_T^2 (\vn \gamma_{ij})^2\right) \right. \label{4}\\
& - 3 c_T^{-2}\dot \Psi^2 + (\vn \Psi)^2 - 2 c_T^{-2} (1+ \alpha_H) \vn \Phi \vn \Psi + \nonumber \\
& \left. c_1 \dot \pi ^2 - c_2 (\vn \pi)^2 + {\rm mixing \ terms} \, \right], \nonumber
\end{align}
where we have defined the GWs speed $c_T^2 = (1+ \epsilon_4)^{-1}$ and the \emph{beyond Horndeski} parameter $\alpha_H = \tilde \epsilon_4 - \epsilon_4$. 
In the Jordan frame there is no direct coupling of $\pi$ to the matter fields, but the scalar-metric mixing terms  schematically indicated in~\eqref{4}, of the type $\vn \Phi \vn \pi$, $\vn \Psi \vn \pi$ and $\dot \Psi \dot \pi$. When $\alpha_H\neq 0$ the higher derivative term $\vn \Psi \vn \dot \pi$ also appears~\cite{GLPV}. 
The explicit form of the last line of~\eqref{4} depend on all the operators of the EFT---\emph{i.e.} also on those omitted in~\eqref{example}---as well as on the time derivatives of  $\epsilon_4$ and $\tilde \epsilon_4$, and is responsible for the rich linear phenomenology of dark energy, in which the $\pi$ fluctuations play a dominant role~\cite{pheno,sawicki,pheno2}. 

However, in the vicinity of a localised matter source the $\pi$ fluctuations become irrelevant because of the screening, and we can thus forget about the third line of~\eqref{4}. 
As long as the on-shell gravitons $\gamma_{ij}$, and the Newtonian potentials $\Phi$ and $\Psi$ can be considered as short-wavelength fluctuations on top of a constant background scalar field gradient, the first two lines of~\eqref{4} can be borrowed from cosmology and applied to general set-ups. This is the case for GWs of wavelengths much shorter than the distance from the source. From the first line of~\eqref{4} we can read off 
$G_{\rm gw} = c_T^2/(8 \pi M^2)$.
For a given (shift-symmetric) theory, the cosmological value of $c_T$ (equivalently, of the EFT parameter $\epsilon_4$) can be calculated as a function of $X = -\dot \phi_0^2$~\cite{GLPV, G32}.
If such a gradient acquires a spatial component $\phi'$---either along the radius from a matter source, or in the direction of our motion w.r.t. the CMB frame---$c_T$ simply transforms as a velocity under a boost of speed $v=\phi'/\dot{\phi}$ and becomes direction-dependent. Along the two principal directions the boosted velocity reads
\be
c_T^{\rm astro} =\ \frac{c_T(X)\pm v}{1\pm c_T(X)v}\, .
\ee

We are left with the \emph{second line} of~\eqref{4}, which can be used to describe the dynamics of the scalar potential gravitons in the Newtonian approximation. However, its applicability to general screened situations is more subtle. Since the Newtonian potentials and the background field $\phi_0$ are generated by the same source, they are of the same typical wavelengths, and the constant gradient approximation for $\phi_0$ is not guaranteed to work. 
By substituting $ g^{\mu \nu}T_{\mu \nu} \simeq - 2 \Phi \rho_m$, one would obtain the relation between the two Newtonian potentials $\gamma_{\rm PPN}\equiv \Psi/\Phi$ and the Newton constant by the Poisson equation:
\begin{align}
\gamma_{\rm PPN} \ = \ \frac{1+\alpha_H}{c_T^2}\, , \qquad G_N  \ = \ \frac{c_T^4}{8 \pi M^2 (1+\alpha_H)^2}\,  .  \label{5}
\end{align}

The study of spherically symmetric configurations in the full beyond Horndeski models confirms that the above always correspond to one available branch of solutions~\cite{Kimura:2011dc,Kobayashi:2014ida}. Theories with terms up to $(\nabla^2\phi)^2$ (\emph{type-4}) show a total of three branches, in two of which the GR result $\gamma_{\rm PPN} \simeq 1$ is recovered inside the Vainshtein radius. However,  for  beyond Horndeski of \emph{type-5} (terms up to terms up to $(\nabla^2\phi)^3$), there appears to be now way to recover the GR value in any of available branches. We would like to emphasize that the branch corresponding to~\eqref{5}, always present, also develops non-linearities inside the Vainshtein radius. A closer inspection of the solutions in~\cite{Kimura:2011dc} shows, however, that for this branch the relevant non-linearities are in the mixed $\pi$-$\Phi$ and $\pi$-$\Psi$ sectors, and not in the self-interactions of the scalar, as it is usually assumed.

The branch recovering $\gamma_{\rm PPN} \simeq 1$, when available, is often taken as the appropriate solution inside virialized objects \citep{Narikawa:2013pjr,Koyama:2015oma}, also because it matches the asymptotically flat solutions in some specific cases \cite{Sbisa:2012zk}. However, which branch applies to realistic scenarios is ultimately selected by the time evolution. The point is to understand, case by case, which solution continuously evolves from the (unique) linear configuration describing a tiny perturbation in the early Universe, and this will depend, in general, on the details of the theory.

\begin{figure}[h]
\includegraphics[width=0.43\textwidth]{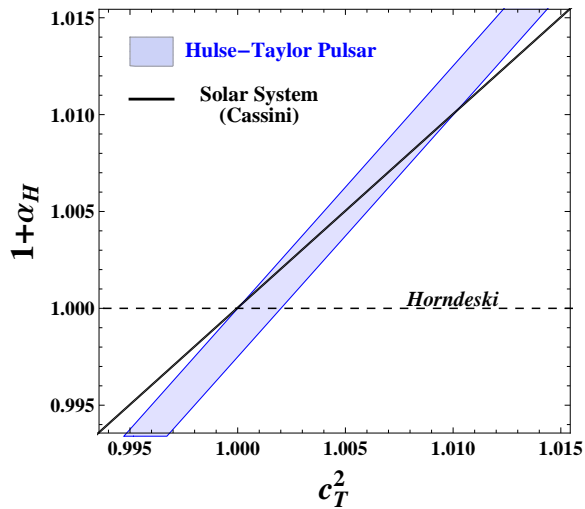}
\caption{Combined constraints in the $(c^2_T ,\alpha_H)$ parameter space for the linear branch of solutions~\eqref{5}. The (tight) bound the Cassini measurement~\cite{cassini} (black curve) and the light-blue stripe corresponding to the Hulse-Taylor pulsar bound obtained from the top panel. Within Horndeski theories, because $G_N = c_T^2 G_{\rm gw}$, the bound~\eqref{new6} turns into a slight preference for superluminal propagation.}
\label{fig2}
\end{figure}

{\bf Observational constraints and discussion -} Within the ``linear branch'' of solutions~\eqref{5}, the cosmological EFT parameters $\epsilon_4$ and $\alpha_H$ are tightly constrained. First, the bound~\eqref{new6} turns into a constraint for the combination of parameters $(1+\alpha_H)^2/c_T$. At the same time, as already noted \emph{e.g.} in~\cite{Kimura:2011dc}, the value of the slip parameter in the linear branch~\eqref{5} is powerfully constrained by the Cassini spacecraft experiment~\cite{cassini}: $\gamma_{\rm PPN}  -1= (2.1\pm2.3)\times 10^{-5}$. This combines with our binary pulsar result  as  in Fig.~\ref{fig2}. 

We would like to stress, however, that beyond the details related to the specific branch of solutions, the bound on the GW speed is very  general: Hulse-Taylor Pulsar's observations constrains $c_T$ at the level of $10^{-2}$, barring remarkable and unlikely cancellations with the (linear and non-linear) physics that determines the orbits of the bound system. Our result applies to all dark energy models in which gravity is modified enough to display a different speed for GWs. Within scalar-tensor theories, in particular, we have considered  galileon 4 and 5 type models and its generalizations and argued that the effect is not screened in general, because it is related to the the persistence of the (cosmological) scalar field gradient even inside conventionally Vainshtein-screened regions.

\textbf{Acknowledgement}:  We acknowledge enlightening conversations with Lam Hui, Tsutomu Kobayashi, Kazuya Koyama, Christian Marinoni, Alberto Nicolis, Louis Perenon, Jeremy Sakstein,  Filippo Vernizzi and Norbert Wex. We especially thank Iggy Sawicki for pointing out a notational inconsistency in a previous version of the paper. This research was funded by grant program of the A*MIDEX Foundation under Contract ANR-11-IDEX-0001-02. HV also thanks support from CNPq and UFES.

\end{document}